\documentclass[a4paper,twocolumn,english,aps,prl,showpacs]{revtex4}
\usepackage[pdftex]{graphicx}
\usepackage{dcolumn,bm,amsmath,natbib,color,amssymb,amsfonts}
\usepackage[T1]{fontenc}
\usepackage[latin1]{inputenc}
\usepackage[normalem]{ulem}
\pdfoutput=1
\newcommand\GGG{$\rm Gd_3Ga_5O_{12}$}
\makeatletter
\usepackage{babel}

\begin{document}
\title{Dispersionless spin waves in Gadolinium Gallium Garnet} 
\author{N.~d'Ambrumenil}
\affiliation{Department of Physics, University of Warwick, Coventry, CV4 7AL, United Kingdom}
\author{O.~A.~Petrenko}
\affiliation{Department of Physics, University of Warwick, Coventry, CV4 7AL, United Kingdom}
\author{H.~Mutka}
\affiliation{
Institut Laue-Langevin, 71 avenue des Martyrs, CS 20156, F-38042 Grenoble Cedex 9, France}
\author{P.~P.~Deen}
\affiliation{European Spallation Source ESS AB, Box 176, SE-221 Lund, Sweden \\ Niels Bohr Institute, University of Copenhagen, Blegdamsvej 17, 2100 Copenhagen, Denmark}

\date{\today}

\begin{abstract}
We report the results of neutron scattering on a powder sample of \GGG\ at high magnetic fields.
We find that in high fields ($B\gtrsim 1.8$~T) the system is not fully polarized, but has a small canting of the moments induced by the dipolar interaction. We show that the degree of canting is accurately predicted by the standard Hamiltonian which includes the dipolar interaction.
The inelastic scattering is dominated at large momentum transfers by a band of almost dispersionless excitations.
We show that these correspond to the spin waves localized on ten site rings, expected for a system described by a nearest neighbor interaction, and that the spectrum at high fields $B\gtrsim 1.8$~T is well-described by a spin wave theory.
The phase for fields $\lesssim 1.8$~T is characterized by an antiferromagnetic Bragg peak at $(210)$ and an incommensurate peak.
\end{abstract}

\maketitle

Gadolinium gallium garnet (GGG) is a frustrated antiferromagnet with a ground state which does not  show long-range order.
The interaction between local ($S=7/2$) moments on the Gd sites is thought to be well described by a short-range exchange interaction together with the long-range dipolar interaction~\cite{KinneyWolf79,Yavors'kii_2006,Yavors'kii_2007}.
While the nature of the low temperature zero-field~\cite{Schiffer_1994,Petrenko_1998,Petrenko_2001} and low-field~\cite{Hov_1980,Schiffer_1995,Petrenko_1999,Dunsiger_2000} properties have attracted much interest, its high field properties have not been so carefully studied~\cite{Petrenko_1999}.
However this high field phase shows remarkable properties.
In particular it has, as lowest lying spin wave excitations, almost dispersionless bands corresponding to excitations localized on 10 site rings~\cite{Ghosh_2008,Henley10,MEZhitomirsky07}.
The magnetic field couples to these excitations via the Zeeman energy and hence contributes to the chemical potential for the (weakly interacting) spin waves.
When the magnetic field is reduced, the effective chemical potential approaches zero. The exact nature of the ground state at these lower fields is unclear but is determined by the interplay between the small dispersion and the interactions between spin waves.

Here we study the high field ferromagnetic (FM) phase of GGG and the transition into an ordered phase with antiferromagnetic (AF) modulations that appears as the applied magnetic field is reduced.
We report the inelastic neutron spectra and the Bragg scattering intensity for a powder sample as a function of applied magnetic field and show that these are in excellent agreement with a spin wave theory valid for fields above $\sim$1.8~T.
In the FM phase we also observe Bragg peaks,  which are forbidden in a fully polarized phase. 
We show that this is a consequence of a canting of the moments induced by the dipolar interaction.
At lower fields other Bragg peaks including those with commensurate and incommensurate wavevectors appear. These are key challenges for any theory of this phase to explain.
\begin{figure}[b]
\includegraphics[width=\columnwidth]{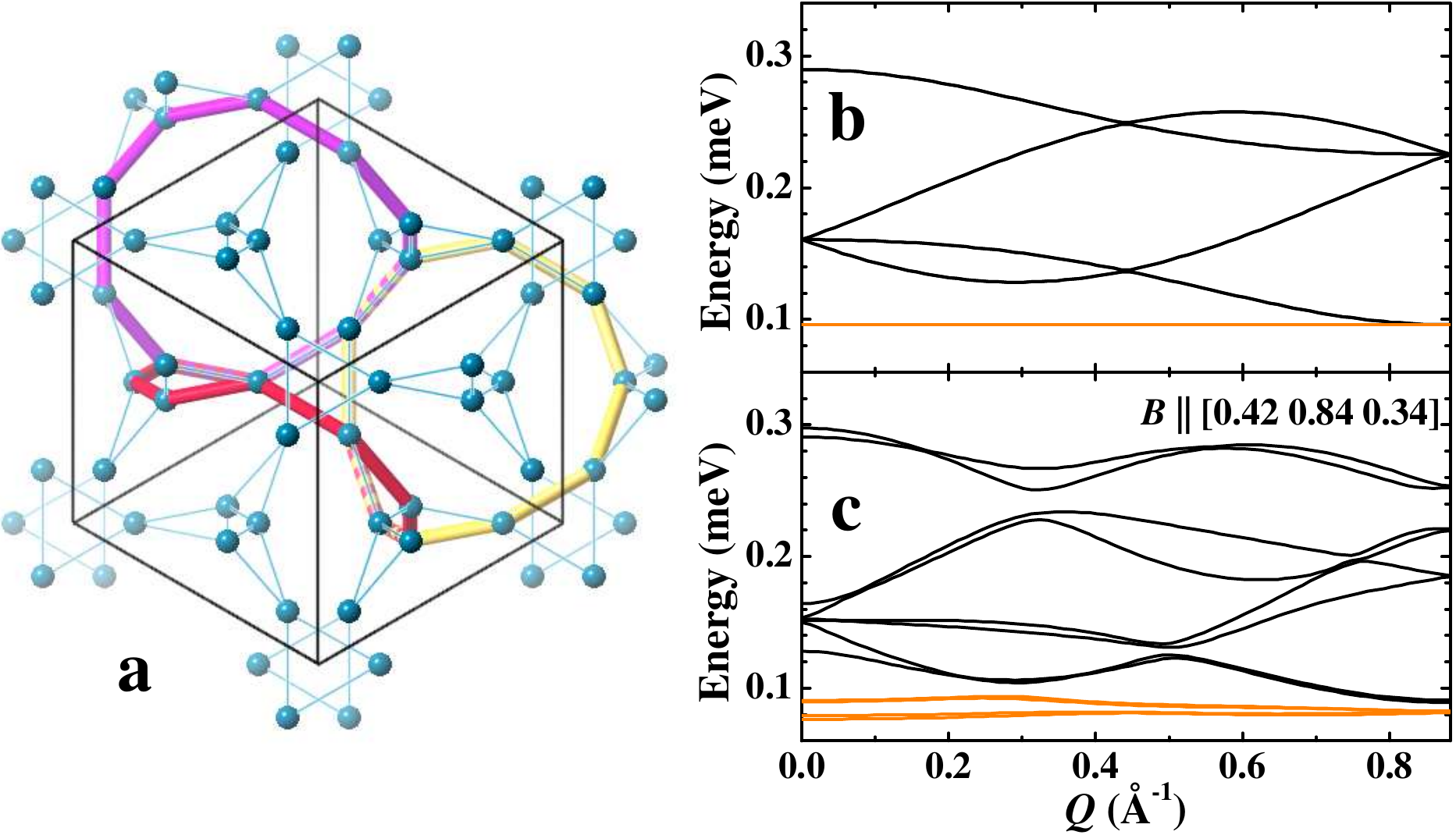}
\caption{\label{fig:Rings} 
(a) Positions of Gd ions in GGG and the cubic unit cell.
The colored lines show rings of Gd ions, on which a spin wave can be localized in a field-induced FM state for a system with nearest neighbor exchange only.
The spin wave bands for $B=2.5$~T for (b) nearest neighbor interactions only and (c) including the dipolar and third nearest neighbor terms (see Eq.~\ref{eq:Hamiltonian}) for a particular field direction. 
The dispersionless (b) and near-dispersionless (c) bands are shown in orange.}
\end{figure}

In GGG, the Gd ions are arranged on a BCC lattice with 24 ions per conventional unit cell, see Fig.~\ref{fig:Rings}(a).
The spin system is well-described by a Hamiltonian  \cite{KinneyWolf79}  with exchange and dipolar interaction (see below, Eq.~\ref{eq:Hamiltonian}).
In a strong magnetic field, the spins align with the applied field and the excitations are the usual spin waves.
However, if only nearest neighbor coupling is included, the lowest excitations show no dispersion at all and the excitations can be localized on 10-site rings, Fig.~\ref{fig:Rings}(a,b).
The dipolar interaction makes the spectrum a function of the relative orientation of the magnetic field and the crystal axes, and introduces some dispersion into the flat bands of the spin waves, Fig.~\ref{fig:Rings}(c).
The physics of the field-induced FM phase, as the magnetic field is lowered and antiferromagnetic modulation sets in, will be dominated by these almost dispersionless modes and the interactions between the spin waves.
\begin{figure}[t]
\includegraphics[width=\columnwidth]{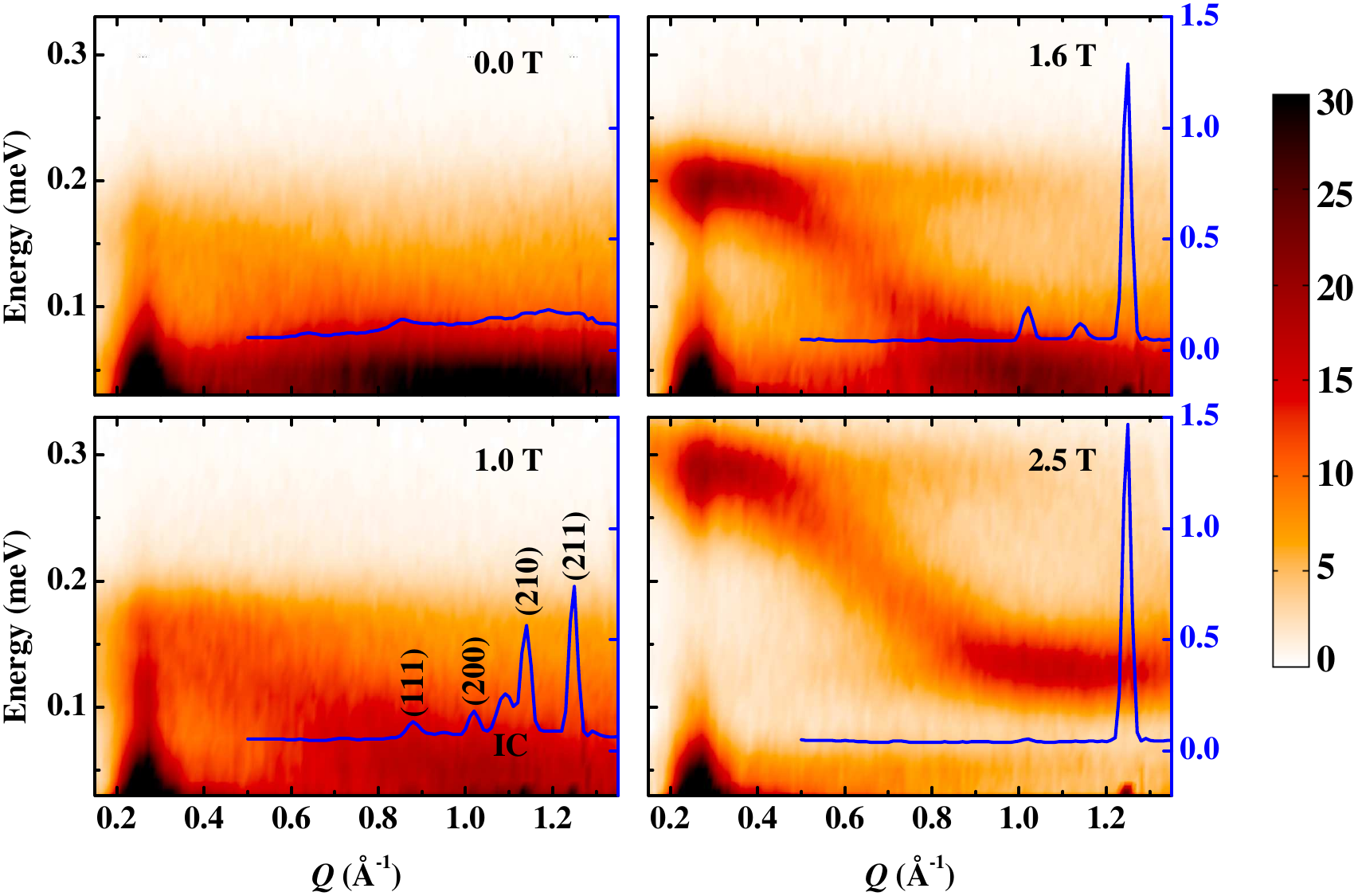}
\caption{\label{fig:spectrum} 
Scattering intensity measured at incident energy $E=1.28$~meV for different applied magnetic fields as a function of momentum transfer $Q$ and energy (left hand axes).
The solid lines show the magnetic Bragg scattering intensities (arbitrary units, right hand axes). 
Two AF peaks are visible at $B=1.6$ and 1.0~T at $(210)$ and at an incommensurate (IC) wavevector.
However, the $(200)$  AF peak, although small, is present even at 2.5~T, as is a small peak at $(110)$.}
\end{figure}

Neutron time-of-flight measurements were performed on a powdered sample using the IN5 direct geometry spectrometer at the ILL with incident energies of $E_i = 1.28$ or 1.94 meV.
The resolution at zero energy transfer (27 or 48 $\mu$eV respectively) was determined for each incident energy using a standard incoherent scatterer.
The field dependence of the  scattering intensity was measured at 0.06~K at fields between 0 and 2.5~T with an identical empty cell at 2~K, 0~T, measured as background.
Higher-temperature data were collected at several fields with different resolution for comparison.
Our sample was the same as in previous investigations \cite{Petrenko_1998,Deen_2010} and contained 99.98\% of the non-absorbing isotope $^{160}$Gd.
The sample was covered with isopropanol 99\% deuterium, that freezes the crystallites into place without any substantial contribution to the scattering.
A cryomagnet used in the experiment has restricted the scattered neutrons to within $\pm 5^\circ$ of the horizontal scattering plane.
Zero-field results, including for the temperature dependence of the magnetic excitations, have been previously reported~\cite{Deen_2010}.
It should be noted that the scattering at $Q\approx 0.25$~\AA$^{-1}$ is an experimental artifact associated with the transmitted beam.

\begin{figure}[t]
\includegraphics[width=\columnwidth]{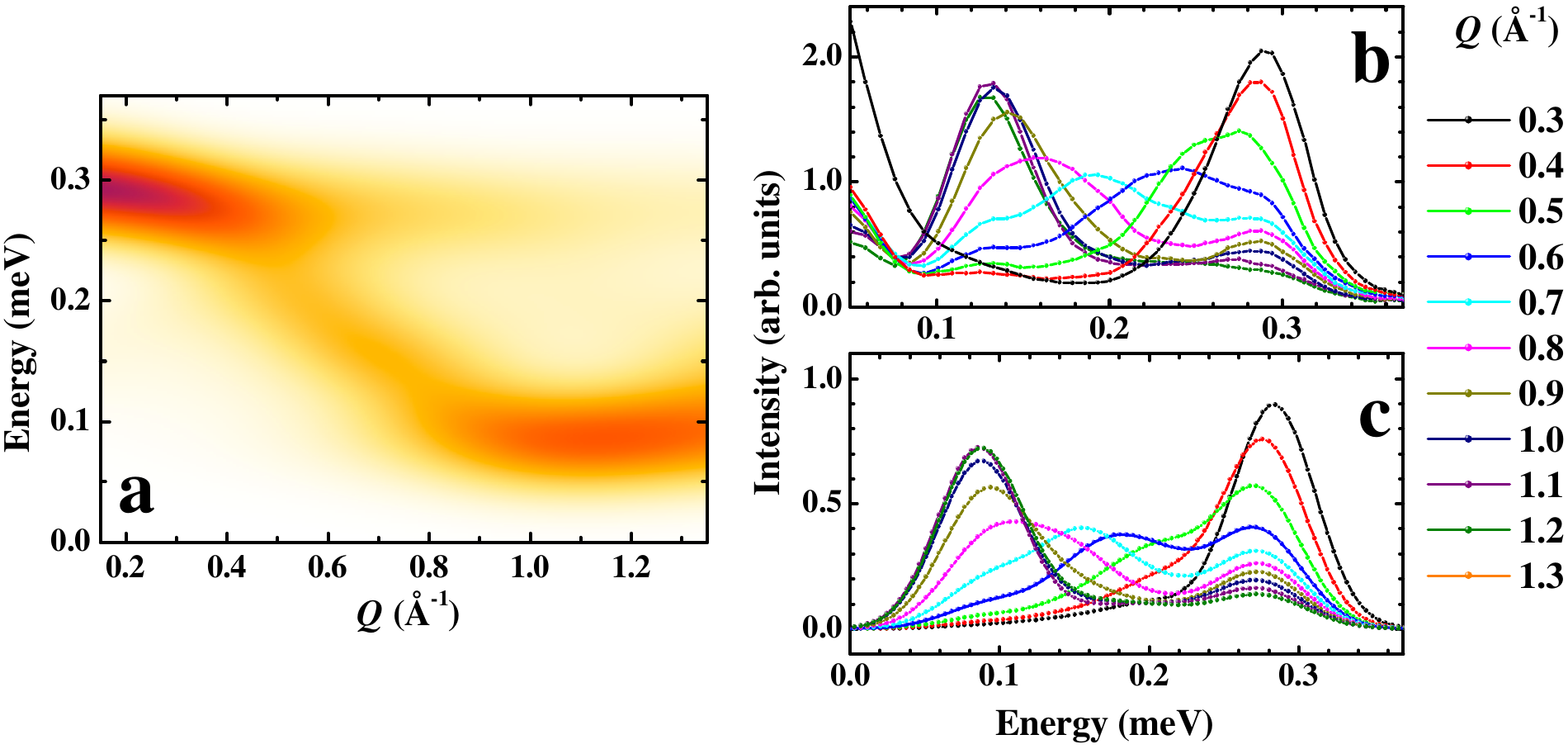}
\caption{\label{fig:Cuts}
(a) Computed scattering intensity for $B=2.5$~T shown as a function of momentum transfer, $Q$, and energy, $E$;
(b) Measured (b) and computed (c) traces for $S(Q,\omega)$ as a function of $\omega$ for a series of momentum transfers $Q$.
The theory assumes an energy resolution of 0.027~meV. }
\end{figure}

Results are shown in Fig.~\ref{fig:spectrum} for four different magnetic fields.
Above about 0.9~\AA$^{-1}$, the scattering is dominated by flat dispersionless bands.
The results are consistent with a spin wave model of the excitations described below, see Fig.~\ref{fig:Cuts}a).
The spectra at 2.5 and 1.6~T have the same shape, with the lower field results shifted down in energy.
To a first approximation, the magnetic field acts to determine the band positions in the spin wave model.   

In Fig.~\ref{fig:Cuts}(b), we show the neutron scattering intensity as a function of $\omega$ for a series of values of $Q$ for a system at $B=2.5$~T and, in  Fig.~\ref{fig:Cuts}(c), the corresponding predictions of our spin wave model.
We see very good qualitative agreement between the measured spectra and the predictions based on a  spin wave picture.
The discrepancies are largely associated with a slightly larger bandwidth in theory than in experiment and the absolute positioning of the spectra.
These discrepancies can be made to disappear if the parameters in the model are altered slightly from the values  suggested in~\cite{KinneyWolf79}.
However, there are also a number of effects in the model which need theoretical exploration before using our calculations to refine the values for the model parameters.
These include the development of the Ewald method to describe matrix elements as well as energies and a treatment of spin non-conserving terms.

\begin{figure}[t]
\includegraphics[width=\columnwidth]{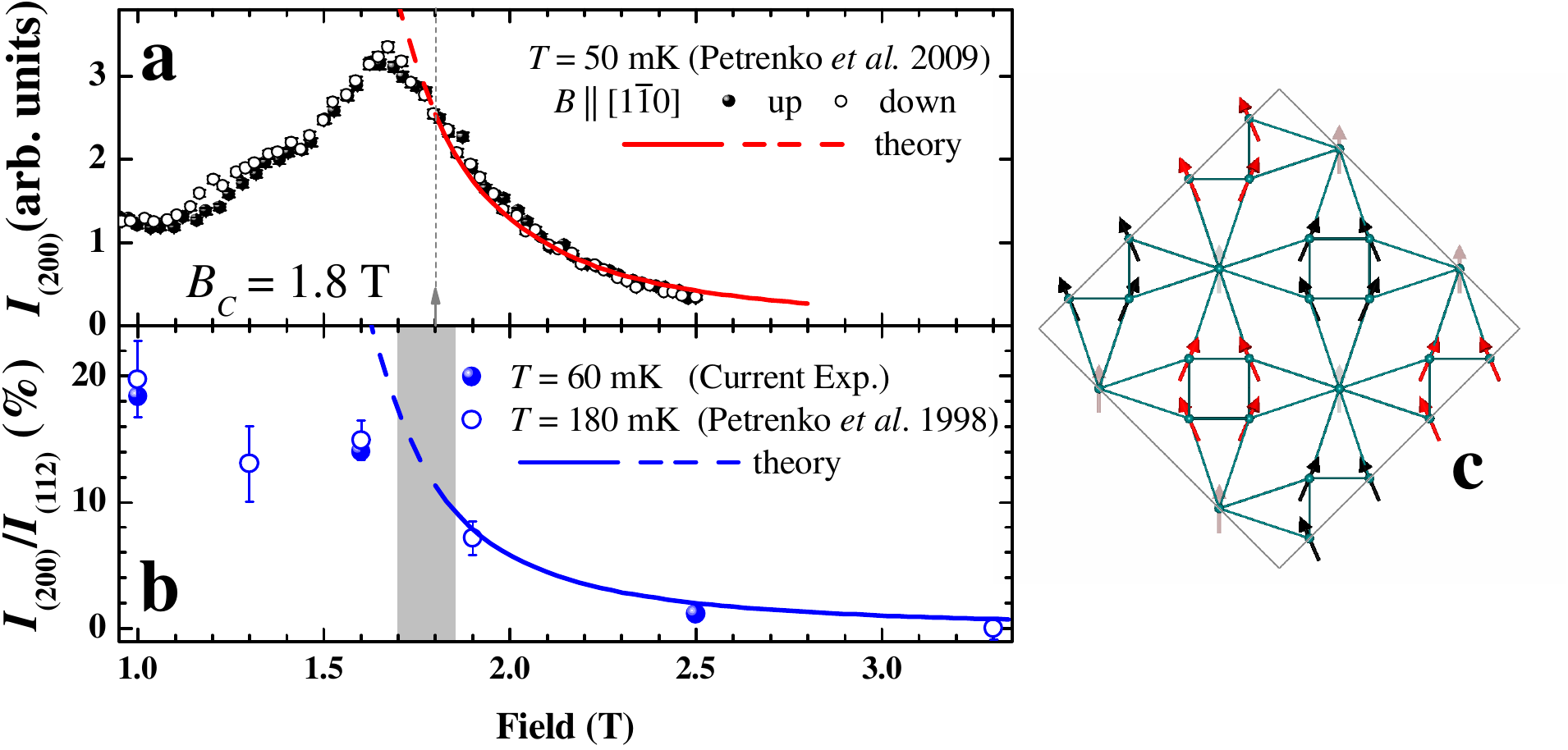}
\caption{\label{fig:Structure} 
Observed and predicted field dependence of the intensity of the $(200)$ magnetic Bragg reflections for: (a) the single crystal studied in Ref.~\onlinecite{Petrenko_2009}, and (b) the powder sample taken at 60~mK (current experiment) and 180~mK previous data \cite{Petrenko_1998}.
For the powder sample we normalise the reflection strength to that of the $(211)$ reflection.
(c) The positions of the Gd ions in GGG.
The red and black arrows show the canting away from the applied field on the two sub-lattices of the magnetic moments induced by the dipolar interaction at high fields ($B\gtrsim 1.8$~T) for the case of {\bf B} along $[110]$. 
The grey arrows show moments aligned with the magnetic field.}
\end{figure}

The field dependence of the intensity of magnetic Bragg peaks is shown as solid lines in Fig.~\ref{fig:spectrum}.
The phase at 1.6 and 1.0~T is characterised by Bragg reflections at an incommensurate wavevector and at $(210)$, which are absent at 2.5~T. However, a $(200)$ reflection, and a weaker one at $(110)$, are both present at 2.5~T.
The strength of the $(200)$ reflection is shown as a percentage of the $(210)$ ferromagnetic reflection in Fig.~\ref{fig:Structure}(b) together with data taken on the same sample  in a different experiment.
The $(200)$ reflection is forbidden for a fully polarised state in the GGG structure, yet there is a clear but weak signal well above the transition fields which are indicated as a shaded region in Fig.~\ref{fig:Structure}(b).
The ill-defined transition region for the powder sample is estimated from previous neutron scattering~\cite{Petrenko_1998} and susceptibility measurements~\cite{Hov_1980} made on single crystals.
These show that the transition field is dependent on the relative orientation of the field with respect to the crystal axes. It is also consistent with our model, as the transition field should be correlated with the field, at which the minimum of the spin wave band approaches zero, and  this varies with the relative orientation of the field and crystal axes.  

We have modeled the spectra using the standard spin model for GGG which assumes that the local moments on each site have $S=7/2$ ($g=2$) and are described by a Hamiltonian including exchange and  dipolar terms~\cite{KinneyWolf79,Yavors'kii_2007}:
\begin{eqnarray}
H & = & \sum_{ j\alpha,l\beta} J_{ j\alpha,l\beta} {\bf S}_{j\alpha} \cdot {\bf S}_{l\beta}-g\mu_B B \sum_{j\alpha} S^z_{l\alpha} + \label{eq:Hamiltonian} \\
& & D \sum_{j\alpha,l\beta} \left( \frac{{\bf S}_{j\alpha} \cdot {\bf S}_{l\beta} - 3 ( {\bf S}_{j\alpha}\cdot \hat{\bf r}_{j\alpha l\beta})( {\bf S}_{l\beta}\cdot \hat{\bf r}_{j\alpha
  l\beta})}{r_{j\alpha,l\beta}^3}\right). \nonumber
\end{eqnarray}
The indices $l$ and $\alpha$ identify the unit cell and the twelve Gd ions in the primitive cell respectively.
The nearest neighbor and third nearest neigbor exchange interactions are $J_1=0.107$~K and $J_3=0.013$~K with all others zero~\cite{KinneyWolf79}.
(A value of $J_2=-0.003$~K is quoted but it has no observable effect on our results.)
The dipolar interaction strength $D r_{nn}^3=0.0457$~K with $r_{nn}$ the nearest neighbor distance.
The vectors $\hat{\bf r}_{j\alpha l\beta}$ denote unit vectors along the vectors joining site $j\alpha$ to site $l\beta$.

In the presence of a large magnetic field, the magnetic moments align with the local field on each site. 
The effect of the dipolar interaction is to add components to the local field in the direction $\hat{\bf r}_{j\alpha l\beta}$ so that the local field is canted with respect to the applied field.
We have computed the direction and degree of canting as a function applied field.
For a given orientation of ${\bf B}$ with respect to the crystal axes, we take the expression for the total ground state energy, given by the Ewald sum over the dipolar interaction for a periodic array of moments on each site in the cubic lattice~\cite{WangHolm01,Leeuwetal80}, and identify the sum of all terms projecting onto the spin ${\bf S}_{l\alpha}$ as the local field.
We align the moments on the site $l\alpha$ with this field and iterate to self-consistency.
We compute the Bragg scattering strength for the resulting spin order \cite{JensenMackintosh91}.

As an example of the modulations induced in the ferromagnetic phase, we show the direction for each site for ${\bf B} \parallel [110]$ in Fig.~\ref{fig:Structure}(c).
In Fig.~\ref{fig:Structure}(a), we show previously unexplained data \cite{Petrenko_2009} taken on a single crystal in a field applied parallel to $[1\bar{1}0]$.  
The theoretical result accounts well for the magnetic field dependence of the $(200)$ Bragg intensity for fields above the transition field in this case of 1.8~T. 

For the powder sample we average the Bragg reflection strength over all orientations of ${\bf B}$ using the oblique array algorithm~\cite{HannayNye04}.
The result for a powder sample is shown in Fig.~\ref{fig:Structure}(b) as a percentage of the ferromagnetic $(211)$ reflection strength.
Above the transition into the FM state (indicated to be fields between 1.7 and 1.85~T), the computed  strength of the weak $(200)$ reflection is compared to the results of the current experiment together with those taken on the same sample at a different spectrometer.
It is clear from Fig.~\ref{fig:Structure} that a modulation induced by the dipolar fields is entirely consistent with the weak $(200)$ reflection observed both in the powder sample and, earlier, in a single crystal. We find similar consistency between theory and observation for the (much weaker) $(110)$ reflection.

The magnon dispersion (Fig.~\ref{fig:Rings}) and spectra (Fig.~\ref{fig:Cuts}) are computed using the Holstein-Primakoff transformation:
\begin{equation}
S^z_{l \alpha} = s -a^\dagger_{l \alpha} a_{l \alpha}, \hspace{0.08in}
S^\dagger_{l \alpha} = \sqrt{2s} \left( 1 -
\frac{a^\dagger_{l \alpha} a_{l \alpha} }{2s}
 \right)^{1/2} a_{l \alpha}.
\label{eq:HPTrans}
\end{equation}
Here the $a^\dagger_{l \alpha}$ satisfy bosonic commutation relations.
Although the Ewald method leads to converged ground state energies, we have not generalised this to the full Hamiltonian and, in particular, to the computation of the matrix elements of the interaction between the neutron and local moment required to compute the inelastic scattering spectra.
Instead we have  included up to seventh nearest neighbors, although, once the experimental resolution is included, there is little observable difference to results obtained by including only nearest neighbor terms.
After introducing Bloch sums,  the resulting $12\times12$ Hamiltonian for a single spin wave introduced into the ferromagnetic states is diagonalized to give 12 separate bands for each crystal momentum, $Q$, which are labeled by $\lambda$.
The scattering intensities shown in Fig.~\ref{fig:Cuts} are computed using the amplitudes of the eigenvectors, $e^\lambda _\alpha (\bf Q)$, via  $| \sum_{\lambda, \alpha}  v(Q) (e_\alpha^\lambda(Q))^* e^{i\bf Q \cdot \alpha} |^2$.
Here $v(Q) \sim \boldsymbol{\mu} . (\hat{\bf Q}\wedge {\bf S} ) \wedge \hat{\bf Q}$ accounts for the interaction between the neutron moment, $\boldsymbol{\mu}$, and the Gd spin~\cite{JensenMackintosh91}, and $\hat{\bf Q}$ is the unit vector parallel to ${\bf Q}$. 

When computing the inelastic spectra, we assume that the spins align parallel with the applied field and take account only of the terms which conserve total spin:
\begin{eqnarray}
 ( {\bf S}_{j\alpha}\cdot \hat{\bf r}_{j\alpha l\beta})( {\bf S}_{l\beta}\cdot \hat{\bf r}_{j\alpha l\beta} ) 
\rightarrow  S _{j\alpha}^z  \left(\hat{r}_{j\alpha l\beta}^z\right)^2 S _{l \beta}^z + \nonumber  \\ 
 \frac{1}{4} \hat{r}_{j\alpha l\beta}^- \hat{r}_{j\alpha l\beta}^+
\left( S _{j\alpha}^+  S _{l \beta}^- + \mbox{h.c.}\right).
\label{eq:spin-conserving} 
\end{eqnarray}
We average over all orientations of the magnetic field with respect to the crystal axes~\cite{HannayNye04}.
We also average over the possible directions for $Q$ with respect to the magnetic field (experimentally the scattering wave-vector is within $\pm 5^\circ$ of being perpendicular to $B$). 
We do not include the effect of the linear terms associated with the canting of the local field away from the applied field at each site described above, although these could be accounted for by reducing the hopping between sites by the overlap between spin directions~\cite{WalkerWalstedt}.
However, at 2.5~T, this overlap is always closer to one than 0.96 and the effects of it deviating from one are smaller than the experimental resolution. 
It would also be possible to take account of the spin non-conserving terms of the type $a_i^\dagger a_j^\dagger$ via a Bogoliubov transformation. 

We show the scattering intensity as a function of frequency for a series of $|Q|$ together with theoretical predictions based on the spin wave picture in Fig.~\ref{fig:Cuts}(b). The correspondence between the theory and experiment is good.
We have worked with the Hamiltonian~(\ref{eq:Hamiltonian}) and not changed the values of the parameters from those estimated from previous zero-field studies of magnetisation and heat capacity \cite{KinneyWolf79}.
In future, a systematic study of the Bragg scattering and a complete theory of the spin wave Hamiltonian (generalising the Ewald method to matrix elements and treating the linear and non-spin conserving terms in the Hamiltonian) should allow a more direct approach to the estimation of the parameters in the model.

The most striking feature of the spectra are the near dispersionless bands apparent in the inelastic spectra for $|Q|\gtrsim 0.9$~\AA$^{-1}$.
Their origin can be seen in the spin wave bands in the FM phase computed without taking account of the dipolar interaction shown in Fig.~\ref{fig:Rings}(a), which shows that these bands are flat across the whole Brillouin zone.
The dipolar interaction introduces some dispersion and, for some alignments of the magnetic field, lifts the degeneracy, which must exist at one point in the Brillouin zone for completely localized excitations~\cite{BergmanBalents08}.
The transition out of the ferromagnetic state into the state with well-developed commensurate $(210)$ and incommensurate peaks in the Bragg scattering, as the field is reduced, is not likely to be of the soft mode type given that there is no clear minimum in the almost flat bands. 

The nature of the state below the transition will be determined by the competing effects of  interactions between the (bosonic) spin waves and the small dispersion introduced by the dipolar interaction into the almost flat bands. 
Two other Gd garnets, $\rm Gd_3Te_2Li_3O_{12}$ (GTLG) and $\rm Gd_3Al_5O_{12}$ (GAG), have the same structure and are described by the same model.
However the ratio $J_1/D$ is estimated to be around 18\% larger in GTLG than in GGG and 25\% larger in GAG \cite{Quilliam_2013}.
With three different values of $J_1/D$ and, in single crystals, with the ability to vary the dispersion relation by varying the relative orientation of the applied field and crystal axes, studies of the three Gd-based garnets should help to establish the nature of the transition and allow exploration of the physics of interacting (nearly-)dispersionless bosons. 

We thank John Chalker for support and guidance.

\bibliographystyle{apsrev4-1}

\bibliography{spinwavesinGGG}
\end{document}